%% file: gc-template.tex
\documentclass[fleqn,11pt]{article}

\usepackage{amsfonts,amsmath,amssymb}
\usepackage{latexsym}

\input preamble.tex

\begin{document}
\twocolumn[
\jnumber{issue}{year}

\Title{On the Accelerated Expansion of the Universe}
	   		
\Author{Naman Kumar\au{1}}
 	     {School of Basic Sciences, Indian Institute of Technology, Bhubaneswar, Odisha, India, 752050}  	     
 	     
\bigskip

\Abstract
	{If we look from a quantum perspective, the most natural way in which the universe can be created is in entangled pairs whose time flow is oppositely related. This suggests the idea of the creation of a universe-antiuniverse pair. Assuming the validity of this hypothesis, in this paper, we show that the universe expands in an accelerated manner. The same reasoning holds for the anti-universe as well. This idea does not require any form of dark energy as used in the standard cosmological model of $\Lambda$CDM or in the modified theories of gravity.
}
\medskip

] 
\email 1 {namankumar5954@gmail.com}

\section{Introduction}
The study of type Ia supernova in 1998 by two independent projects, the Supernova Cosmology Project and the High-Z Supernova Search Team, revealed an astonishing result that the universe is not only expanding but it is expanding at an accelerating rate \cite{1}. The accelerated expansion of the universe is thought to have begun since the universe entered its dark-energy-dominated era roughly 5 billion years ago. The currently accepted theory of gravity, General Relativity, accounts for this accelerated expansion by introducing a small and positive value of the cosmological constant $\Lambda$. Alternative explanations, such as quintessence (see \cite{2} for a review), are also hypothesized as the explanation for cosmic acceleration. Some explanations beyond four dimensions also exist, such as the Dvali-Gabadadze-Porrati (DGP) model\cite{3}, which argues that gravity behaves as 5D at large distances while reducing to that of usual 4D gravity at short distances. However, the precise nature of dark energy remains unknown. Alternative explanations are promising but have their own shortcomings, as was argued in \ cite {4} that the self-accelerating solution of the DGP model \ cite {5} contains a ghost mode. Very intriguing arguments are presented in \cite{6,7,8,9} that the universe should exist in pairs with their time oppositely related. In this Letter, we show using recently conjectured Quantum Focusing\cite{10} that if the universe exists as a pair of universe-antiuniverse pairs, it must expand in an accelerated manner. We begin our discussion by reviewing the Quantum Focusing Conjecture (QFC).

\subsection{Quantum Focusing Conjecture}
In this section, we review the formulation of the Quantum Focusing Conjecture (QFC)
 \subsubsection{Generalized Entropy for \\Cauchy-splitting surfaces}
\begin{figure}
\centering
\includegraphics[width=8 cm]{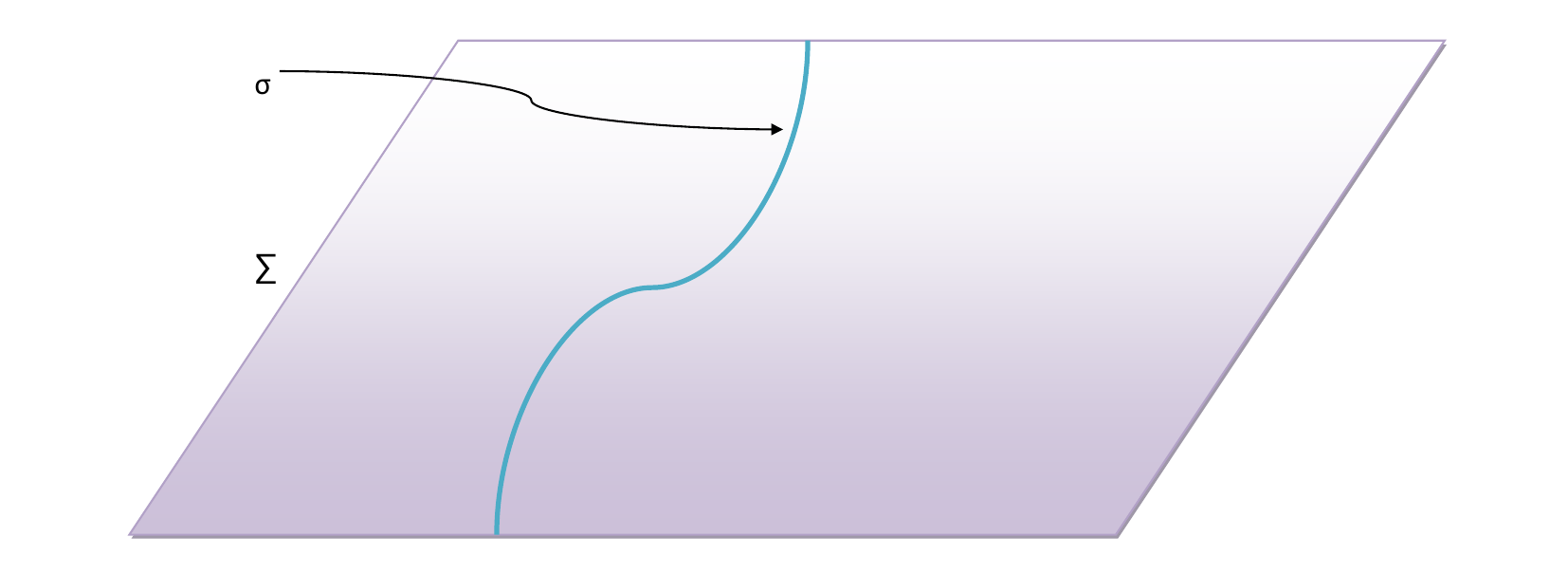}
\caption{\small A Cauchy surface $\Sigma$ is divided into two parts by a surface $\sigma$. $S_{out}$ is defined as entropy restricted to any one side of the splitting surface $\sigma$.}
\label{fig3}
\end{figure}

Generalized entropy was originally defined in \cite{11} in asymptotically flat space as the area A of all black hole horizons, plus the entropy of matter systems outside the black holes:\
\begin{equation}
    S_{gen}=S_{out}+\frac{A}{4G\hbar}+counterterms
\end{equation}\
A rigorous definition of $S_{out}$ can be given as the von Neumann entropy of the quantum state of the exterior of the horizon:
\begin{equation}
    S_{out}=-tr\rho_{out}ln\rho_{out}
\end{equation}
The GSL was introduced to keep the second law of thermodynamics intact when matter entropy is lost in a black hole. Bekenstein conjectured that GSL \cite{11} survives: the area increase of the black hole will compensate for the lost matter entropy so that the generalized entropy will not decrease.
The notion of generalized entropy can be extended beyond the context of causal horizons \cite{10}. Let $\sigma$ be a spacelike codimension-2 surface that splits a Cauchy surface $\Sigma$ into two portions. By picking any one of the two sides of $\sigma$ arbitrarily, we can define an entropy restricted to one side of $\sigma$ as $S_{out}$ (Fig.(\ref{fig3})).
\subsubsection{Quantum Focusing Conjecture}
It conjectures\cite{10} that the quantum expansion $\Theta$, where $\Theta$ is given by:
\begin{equation}
    \Theta=\theta+\frac{4G\hbar}{\mathcal{A}}S'_{out}
\end{equation}
($\theta$ is the classical expansion and $\mathcal{A}$ is the width of null congruence along its generator), cannot increase along any congruence, which is valid for quantum states too:
\begin{equation}
    \frac{d\Theta}{d\lambda}\leq0
\end{equation}
where $\lambda$ is an affine parameter.
The evolution of the expansion $\theta$ along congruence is determined by the Raychaudhuri equation:
\begin{equation}
    \frac{d\theta}{d\lambda}=-\frac{\theta^2}{2}-\sigma_{ab}\sigma^{ab}-R_{ab}k^ak^b:
\end{equation}
where $R_{ab}$ is the Ricci tensor, $\sigma_{ab}$ is the shear, and $k^a$ is the (null) tangent vector to the congruence.
This gives QFC as:
\bearr
0\geq \Theta'=\theta'+\frac{4G\hbar}{\mathcal{A}}(S''_{out}-S'_{out}\theta)
\nnn=-\frac{1}{2}\theta^2-\zeta^2-8\pi G\langle T_{kk}\rangle
\nnn+\frac{4G\hbar}{\mathcal{A}}(S''_{out}-S'_{out}\theta)
\ear
$\zeta$ is shear.
The special choice of congruence for $\theta=\zeta=0$, gives the Quantum Null Energy Condition (QNEC)
\begin{equation}
   \langle T_{kk}\rangle- \frac{\hbar}{2\pi\mathcal{A}}S''_{out}\geq0\label{qnec}
\end{equation}
\subsection{Relative Entropy and Bekenstein Bound}
The Bekenstein bound is given as
\begin{equation}
    S\leq 2\pi ER\label{casbb}
\end{equation}
where $R$ is the radius of the boundary of the system and $E$ is the total energy.
Casini\cite{12} showed that the above bound can be written in the form
\begin{equation}
    S_{V}\leq K_{V}\label{casb}
\end{equation}
where $S_V$ is a localized entropy and $K_V$ is a localized energy.
Using $K=-\log\rho^0_V-\log(tre^{-K})$ and $tr\rho_V=tr\rho^0_V=1$, this can be written as
\begin{equation}
    tr(\rho_V\log\rho_V)-tr(\rho_V\log\rho^0_V)\geq0
\end{equation}
This is simply the statement of the positivity of the relative entropy $S(\rho_V||\rho^0_V)$ between the state $\rho_V$ and the vacuum state $\rho^0_V$, and thus the bound holds. For an object in a half-space that is far from the circumscribing boundary, (\ref{casb}) reduces to (\ref{casbb}). We will use this limit in the next section.
\section{Towards Accelerated Expansion}
\begin{figure}[ht!]
    \centering
    \includegraphics[width=10 cm]{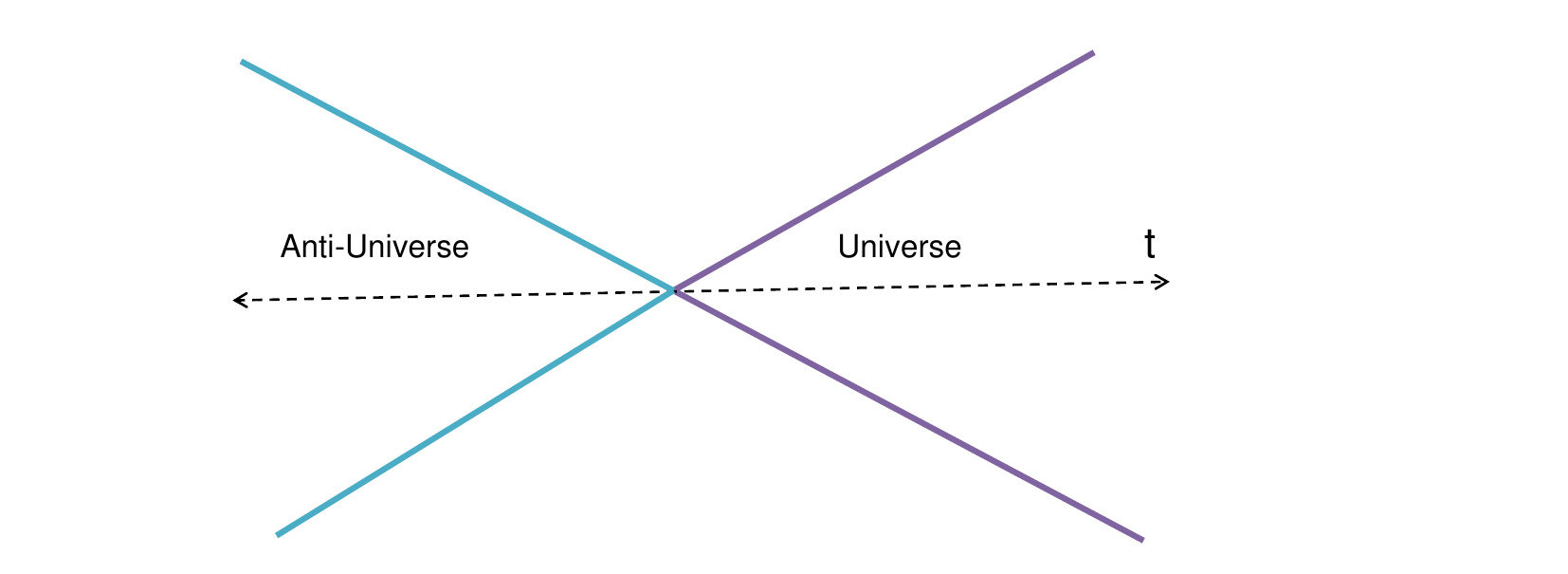}
    \caption{\small A diagram showing an entangled universe-antiuniverse pair. The universe runs in the future($t>0$) while the anti-universe runs in the past($t<0$). The surface at $t=0$ is the entangling surface. We choose the half-space to be the region given by $t>0$ (the universe of the universe-antiuniverse pair). The modular Hamiltonian $K$ is restricted to act on this half-space. Therefore, for this setup, we can apply both Casini's idea and the QNEC.}
    \label{fig2}
\end{figure}

\begin{figure}
    \centering
    \includegraphics[width=10 cm]{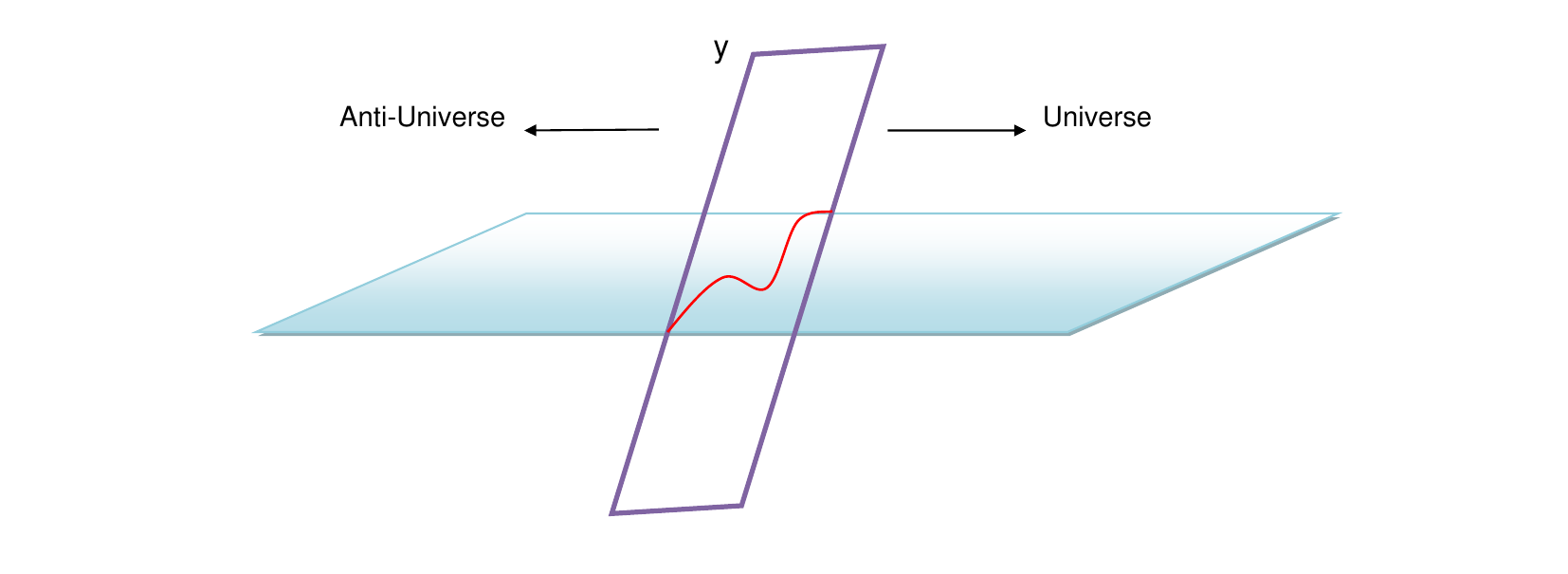}
    \caption{\small A Cauchy surface (horizontal blue surface) is divided into two parts by $\gamma$ (shown by a red curve). $\gamma$ is a section of the null surface $y$ (at $t=0$).}
    \label{fig1}
\end{figure}

A measure of distinguishability between any two states of an arbitrary quantum system is relative entropy. It is a quantity of particular significance in quantum information theory. Its illuminating use in the gravitational context was put by Casini\cite{12}. Consider a null surface $N$ that is split into two parts by a codimension-2 plane $\sigma$. For states restricted to the part of $N$ in the future of $\sigma$ together with a portion of null infinity, the relative entropy is related to the energy-momentum tensor $T_{ab}$ as\cite{10}
\begin{equation}
    S''(\rho||\rho^{(0)})=\frac{2\pi\mathcal{A}}{\hbar}\langle T_{kk}\rangle-S''_{out}\label{rel}
\end{equation}
where $\rho^{(0)}$ is the vacuum state and $\rho$ is some other state. $T_{kk}=T_{ab}k^ak^b$ where $k^a$ and $k^b$ are null vectors. (\ref{rel}) essentially tells that the second derivative of the relative entropy with one of the states being the vacuum state is non-negative. This can be easily verified by noticing that the right-hand side of (\ref{rel}) is simply the statement of QNEC(\ref{qnec}).
The relative entropy, in turn, can be written as the statement of the Bekenstein Bound as
\begin{equation}
   S(\rho||\rho^{(0)})=K_V-S_V\label{rel1}
\end{equation}
where $K_V=tr(K\rho_V)-tr(K\rho^{(0)}_V)$ with $\rho_V$ and $\rho_V^{(0)}$ being reduced density matrices defined on half-spatial plane $V$ (whose causal development is the Rindler wedge). More precisely, $V$ is the cut of the Cauchy surface $t=0$ defined by $z\geq0$. $S_V=S_{out}-S_{out}^{(0)}$ is the vacuum subtracted entropy, where, $S_{out}=S(\rho_V)$ and $S_{out}^{(0)}=S(\rho_V^{(0)})$. $K$ is the modular Hamiltonian that acts on the Right Rindler wedge. We now apply this result to the universe as a whole. Since the arguments presented hold locally (for a Rindler horizon), the extension to the entire universe requires a global existence of a causal horizon. This is similar to the existence of the Unruh effect for a local Rindler horizon to that of the Hawking effect for a global event horizon of a black hole. This global extension can be achieved by considering a universe-antiuniverse pair, in which case the surface at $t=0$ naturally acts as a causal horizon (Fig.\ref{fig2}). The universe runs in the future($t>0$) while the anti-universe runs in the past($t<0$). The universe-antiuniverse pair is classically disconnected but quantum-mechanically entangle,d with the surface at $t=0$ being the entangling surface. In this case, we can define an entanglement entropy $S(\rho_V)$ and vacuum entanglement entropy $S(\rho_V^{(0)})$ where $\rho_V$ is the reduced density matrix of the universe defined on the cut of Cauchy surface by $\gamma$ (see Fig.(\ref{fig1})) obtained by tracing out the degrees of freedom of the anti-universe. The modular Hamiltonian $K$ acts on the half-space defined by the region $t>0$ (universe of the anti-universe pair). Thus, the presence of a universe-antiuniverse pair is required to define the entanglement entropies. Moreover, the pair constitutes the whole space with the universe causally disconnected from the anti-universe, and lies in the half-space where $K$ is restricted to act. These two key insights allow the global extension of the above results to the entire universe. Note that the entropy $S_{out}$ refers to the entropy on one side of the Cauchy surface, but by unitary time evolution is equivalent to states restricted to a null surface $N$\cite{10,13}. So, consider a Cauchy surface which is split into two parts by a spatial surface $\gamma$ which lies on the null surface $y$ at $t=0$ (see Fig.(\ref{fig1})). Therefore, states on the Cauchy surface are unitarily equivalent to the states restricted to the null surface $y$. In this case, we make use of the fact that the vacuum entanglement entropy on null cuts is stationary (\cite{10,13,14,15}). Therefore, we can now apply (\ref{rel}) and (\ref{rel1}) to the universe as a whole. Using (\ref{rel1}), (\ref{rel}) becomes ($S''_{out}$ cancels on both sides and $S^{''(0)}_{out}=0$ since the vacuum entropy is stationary)
\begin{equation}
    K_V''=\frac{2\pi\mathcal{A}}{\hbar}\langle T_{kk}\rangle\label{kv} 
\end{equation}
Applying now (\ref{kv}) to the half-space of the universe, we first make the assumption that the universe respects the Null Energy Condition(NEC) so that $\langle T_{kk}\rangle>0$. Thus, (\ref{kv}) becomes an inequality as
\begin{equation}
     K''_V>0\label{kv1}
\end{equation}
The affine parameter $\lambda$ can now be taken to be the cosmic time $t$, and the total energy $E$ of the universe is constant.
Considering the circumscribing radius of the universe to be $R$ with $R=\infty$ (can be thought of as an infinite imaginary sphere of radius $R$ circumscribing the universe).
Then, for this region in the half-space, we have (following Casini) $K_V\equiv 2\pi ER$. Therefore, we can finally write (\ref{kv1}) as
\begin{equation}
\ddot R>0
\end{equation}
This equation tells us that the second derivative of the circumscribing radius is positive, which means that the universe is expanding in an accelerated manner, all by itself! (provided the universe respects NEC) and does not require any hypothetical form of energy such as dark energy (which is the cosmological constant of $\Lambda$CDM). The same reasoning can be applied to the anti-universe by considering it as the half-space defined by the region $t<0$, so that it also expands in an accelerated manner. Therefore, in general, both the universe and the anti-universe are expanding, which is consistent with \ cite {8}.
\\
\section{Conclusion}
 In this Letter, we showed that if the universe exists as a universe-antiuniverse pair, then it expands in an accelerated manner. The same reasoning can be extended to the anti-universe if we take it to be a half-space defined by the region $t<0$. An important condition that is required to hold is the Null Energy Condition(NEC). A familiar statement exists in the study of black holes, where the horizon area is a non-decreasing function of time, which is the famous Hawking area theorem. It is interesting to note that Hawking's area theorem also requires the Null Energy Condition (NEC) to hold. This suggests that the study of causal horizons reveals a deeper understanding of the universe. If our approach is correct, then we need no dark energy to explain the present acceleration of the universe, and we may need to modify notions of our standard cosmology.

\Acknow{We would like to thank the anonymous referee whose comments greatly improved the clarity of the manuscript.
}

\Funding{There are no funding sources to disclose.
}

\clearpage
\onecolumn
\twocolumn[
\begin{@twocolumnfalse}
\begin{center}
{\Large\bfseries Addendum: Clarification of the passage from Eqs.~(14) to (15)\par}
\end{center}
\vspace{0.5em}
\noindent\textit{Added to the arXiv version in October 2026.}---The published version of this Letter is unchanged. This addendum makes explicit several intermediate steps and assumptions entering the passage from Eq.~(14), $K_V''>0$, to Eq.~(15), $\ddot R>0$. These points concern the null-cut/Cauchy-surface correspondence, the large-radius limit, the modular Hamiltonian, vacuum QNEC saturation, the role of the null-energy condition, and the constancy of the matter energy used in the final reduction. We thank Michael VanBeuzekom for questions that motivated this clarification.
\vspace{1em}
\end{@twocolumnfalse}
]

\subsection*{A. Null-cut variation and cosmological time}
The entropy $S_{out}$ in the main text is defined on one side of a Cauchy surface, while the same state data can be represented on the null surface by unitary evolution, as already used in the derivation preceding Eq.~(13) and discussed in Refs.~\cite{10,13,14,15}. Let $\Sigma_t$ denote the cosmological Cauchy foliation and let $\mathcal N$ denote the null surface used in the QNEC description. At each cosmological time, the reduced state associated with the relevant Cauchy region has a corresponding representation by a cut $\gamma(\lambda)\subset\mathcal N$. The null cut is therefore not the physical trajectory of the boundary of the comoving region; it is the null-surface representation of the same reduced state data. We denote this unitary correspondence by
\begin{equation}
\Sigma_t\ \overset{U}{\longleftrightarrow}\ \gamma(\lambda)\subset\mathcal N.
\end{equation}
The parameter $\lambda$ labels the cuts in the null representation, while $t$ labels the corresponding cosmological Cauchy evolution. The identification used in the Letter is the common parametrization of corresponding members of these two unitarily equivalent families,
\begin{equation}
\gamma(\lambda=t)\longleftrightarrow\Sigma_t,
\label{add:lambda_t}
\end{equation}
so that the null-cut variation is represented by the cosmological-time variation of the corresponding Cauchy data. With this choice of common label,
\begin{equation}
\frac{d}{d\lambda}=\frac{d}{dt},
\qquad
\frac{d^2}{d\lambda^2}=\frac{d^2}{dt^2}.
\end{equation}
Thus the step does not require identifying cosmic time with the affine parameter of an FLRW null geodesic. It uses the null surface as the unitarily equivalent representation of the time-labelled Cauchy evolution.

\subsection*{B. Relation of $R(t)$ to the scale factor and the large-radius limit}
The quantity $R(t)$ is the physical radius of a large circumscribing comoving region, not a physical edge of the Universe. For a fixed comoving radius $\chi_0$ in an FLRW geometry,
\begin{equation}
R(t)=a(t)S_k(\chi_0),
\end{equation}
where $S_k(\chi)=\sin\chi$, $\chi$, or $\sinh\chi$ for positive, zero, or negative spatial curvature. Hence
\begin{equation}
\ddot R(t)=S_k(\chi_0)\,\ddot a(t),
\end{equation}
and for a nondegenerate region $S_k(\chi_0)>0$ the signs of $\ddot R$ and $\ddot a$ agree.

The notation $R=\infty$ in the published text was intended as shorthand for taking the circumscribing surface arbitrarily far away so that the construction represents the whole Universe. A regulated statement is obtained by introducing finite radii $R_L(t)$ for increasingly large comoving regions, taking the derivatives at finite $L$, and only afterwards taking $L\to\infty$ when the relevant limit exists. No derivative of the mathematical constant $\infty$ is involved.

\subsection*{C. Modular Hamiltonian, vacuum saturation, and the role of NEC}
The exact starting point is the modular Hamiltonian rather than the expression $2\pi ER$ by itself. Casini's relative-entropy formulation gives the vacuum-subtracted modular energy exactly \cite{12}. More directly, Koeller, Leichenauer, Levine, and Shahbazi-Moghaddam derive for a smooth null cut $v=V(y)$ the vacuum-subtracted modular Hamiltonian \cite{16}
\begin{equation}
\Delta K_\sigma
=\frac{2\pi}{\hbar}\int d^{d-2}y
\int_{V(y)}^{\infty}dv\,\bigl(v-V(y)\bigr)T_{vv}.
\label{add:koeller}
\end{equation}
Their derivation assumes QNEC together with its saturation in the reference vacuum. The saturation input is already present in the original construction: the vacuum entropy on the null cuts is taken to be stationary,
\begin{equation}
S_{out}^{(0)\prime\prime}=0.
\end{equation}
For the reference vacuum, $\langle T_{kk}\rangle_0=0$, and therefore
\begin{equation}
\frac{2\pi}{\hbar}\langle T_{kk}\rangle_0
-S_{out}^{(0)\prime\prime}=0,
\end{equation}
which is precisely vacuum QNEC saturation in the null-plane construction. Thus saturation is not an additional assumption introduced only in this addendum; it is the stationarity condition already used in obtaining Eq.~(13).

With QNEC and vacuum saturation, the second null-cut variation reduces to
\begin{equation}
K_V''=\frac{2\pi\mathcal A}{\hbar}\,\langle T_{kk}\rangle.
\label{add:qnecstress}
\end{equation}
Here $\mathcal A$ is the infinitesimal transverse area element associated with the local null deformation in the QNEC construction; it is not the macroscopic area $4\pi R^2$ of the circumscribing comoving ball. For the global reduction used in the Letter, restoring the explicit quantum normalization gives
\begin{equation}
K_V=\frac{2\pi}{\hbar}ER.
\label{add:ker}
\end{equation}
Here $E$ and $R$ have the localized-system meaning used in Casini's QFT formulation of the Bekenstein bound \cite{12}: $E$ is the energy of the localized subsystem and $R$ its characteristic circumscribing size. In the present cosmological realization that subsystem is the chosen large comoving ball. The null-cut representation supplies the modular variation; it does not redefine $E$ as an independently varying ``energy beyond the cut.''
When $E$ is constant and the common cut/Cauchy parameterization \eqref{add:lambda_t} is used,
\begin{equation}
K_V''=\frac{2\pi E}{\hbar}\,\ddot R.
\end{equation}
Combining this with Eq.~\eqref{add:qnecstress}, the common factor $2\pi/\hbar$ cancels exactly,
\begin{equation}
E\,\ddot R=\mathcal A\,\langle T_{kk}\rangle.
\label{add:classical}
\end{equation}
The final relation has no explicit $\hbar$. At this stage the sign is therefore fixed by the classical null-energy condition. For $E>0$, $\mathcal A>0$, and
\begin{equation}
\langle T_{kk}\rangle>0,
\end{equation}
Eq.~\eqref{add:classical} gives $\ddot R>0$. In this sense QNEC supplies the quantum-information-theoretic relation, vacuum saturation removes the entropy term, and the modular-Hamiltonian substitution removes the remaining explicit quantum normalization; the final sign criterion is the classical NEC.

Koeller et al. also show that the null-cut result extends from flat-space Rindler horizons to cuts of bifurcate Killing horizons in arbitrary spacetimes, provided QNEC and the corresponding vacuum-saturation condition hold \cite{16}. In the universe--antiuniverse construction, the $t=0$ separating horizon between the two branches supplies the relevant causal/entangling horizon. The construction assumes that this surface supplies the analogous bifurcate geometric modular structure, in the same sense that the $t=0$ entangling surface of the left and right Rindler wedges is the bifurcation surface of the boost horizon. This is the global-horizon assumption entering the curved application.

\subsection*{D. Matter-energy conservation}
The energy entering Eq.~\eqref{add:ker} can be identified with the matter energy inside the corresponding fixed comoving regulated region during pressureless matter domination. Local stress-energy conservation gives
\begin{equation}
\dot\rho+3H(\rho+p)=0.
\end{equation}
For dust, $p=0$, so
\begin{equation}
\rho_m\propto a^{-3}.
\end{equation}
The physical volume of a fixed comoving region scales as $V_{\rm phys}\propto a^3$, and consequently
\begin{equation}
E_m=\rho_m V_{\rm phys}=\mathrm{constant}.
\end{equation}
Thus $\dot E_m=\ddot E_m=0$, exactly as required in the differentiation of Eq.~\eqref{add:ker}. This statement concerns the matter energy in the specified comoving region; it is not a claim that an arbitrary notion of total energy is conserved in a general cosmological spacetime.

If one instead allowed a general time-dependent $E(t)$, then
\begin{equation}
\frac{d^2K_V}{dt^2}
=\frac{2\pi}{\hbar}
\left(E\ddot R+2\dot E\dot R+R\ddot E\right),
\end{equation}
and $K_V''>0$ would not by itself determine the sign of $\ddot R$.

\subsection*{E. Matter domination and the origin of acceleration}
The statement is not that an additional dark-energy component is introduced alongside pressureless matter. The matter sector remains the ordinary pressureless matter sector. In the universe--antiuniverse construction, the same matter stress tensor enters the QNEC/modular relation, and Eqs.~\eqref{add:qnecstress} and \eqref{add:ker} give
\begin{equation}
E\ddot R=\mathcal A\langle T_{kk}\rangle.
\end{equation}
Thus positive matter energy satisfying the NEC supplies a positive acceleration within the universe--antiuniverse modular construction, without introducing an independent dark-energy stress tensor.

Standard pressureless FLRW evolution in Einstein gravity contains the usual matter-induced decelerating contribution. A complete embedding of the present mechanism into the background cosmological dynamics would have to incorporate the additional constraint supplied by the universe--antiuniverse modular structure. In such a formulation, the resulting positive contribution may ultimately dominate the usual matter-induced deceleration. Its source nevertheless remains the same matter sector together with the global entanglement/modular structure, rather than a new dark-energy component. Deriving the corresponding complete background equation and determining the epoch at which this contribution becomes dominant are beyond the scope of the present work.

\subsection*{F. Logical summary}
The clarified chain of implications is
\begin{equation}
\begin{aligned}
\text{QNEC + vacuum saturation}
&\Longrightarrow K_V''>0,\\[-1mm]
K_V''&=\frac{2\pi\mathcal A}{\hbar}\langle T_{kk}\rangle.
\end{aligned}
\end{equation}
\begin{equation}
K_V=\frac{2\pi ER}{\hbar}
\Longrightarrow
E\ddot R=\mathcal A\langle T_{kk}\rangle.
\end{equation}
\begin{equation}
\text{NEC}
\Longrightarrow
\ddot R>0\Longleftrightarrow\ddot a>0.
\end{equation}
The second line uses the unitary Cauchy/null correspondence with the common labeling $\lambda=t$, the geometric modular-Hamiltonian/global-horizon specialization, and the conserved positive matter energy of a fixed comoving region in the pressureless matter regime. Equation~(15) should therefore be read as a result within these assumptions, rather than as an unconditional consequence of QNEC alone.

\end{document}

%% file: preamble.tex
\usepackage{amsfonts,amssymb,cite}
\usepackage{graphicx}

\def\noi{\noindent}
\def\jnumber#1#2{\thispagestyle{empty} \noi\unitlength=1mm
    	\begin{picture}(178,10)
            \put(177,15){\llap{\large\it Grav. Cosmol. No.\,#1, #2}}
                    \end{picture}}

\newcommand{\Title}[1]{\noi {{\Large\bf #1}}\\[1ex]}

\newcommand{\Author}[2]{\noi{\bf #1}\\[2ex]\noi{\normalsize\it #2}\\}

\def\au#1{${}^{#1}$}

\newcommand{\Abstract}[1]{\vskip 2mm \begin{center}
        \parbox{16.4cm}{\small\noi #1} \end{center}\medskip}

\def\email#1#2{\footnotetext[#1]{e-mail: #2}\addtocounter{footnote}{1}}

\def\nqq{\hspace*{-2em}}

\usepackage{color}

\def\Acknow#1{\subsection*{Acknowledgments} #1}
\def\Funding#1{\subsection*{Funding} #1}

\def\Jl#1#2{#1 {\bf #2},\ }

\def\ApJ#1 {\Jl{Astroph. J.}{#1}}
\def\CQG#1 {\Jl{Class. Quantum Grav.}{#1}}
\def\DAN#1 {\Jl{Dokl. AN SSSR}{#1}}
\def\GC#1 {\Jl{Grav. Cosmol.}{#1}}
\def\GRG#1 {\Jl{Gen. Rel. Grav.}{#1}}
\def\IJMPD#1 {\Jl{Int. J. Mod. Phys. D}{#1}}
\def\JETF#1 {\Jl{Zh. Eksp. Teor. Fiz.}{#1}}
\def\JETP#1 {\Jl{Sov. Phys. JETP}{#1}}
\def\JHEP#1 {\Jl{JHEP}{#1}}
\def\JMP#1 {\Jl{J. Math. Phys.}{#1}}
\def\NPB#1 {\Jl{Nucl. Phys. B}{#1}}
\def\NP#1 {\Jl{Nucl. Phys.}{#1}}
\def\PLA#1 {\Jl{Phys. Lett. A}{#1}}
\def\PLB#1 {\Jl{Phys. Lett. B}{#1}}
\def\PRD#1 {\Jl{Phys. Rev. D}{#1}}
\def\PRL#1 {\Jl{Phys. Rev. Lett.}{#1}}

\def\lal{&&\nqq {}}

\def\beq{\begin{equation}}
\def\eeq{\end{equation}}
\def\bear{\begin{eqnarray}}
\def\bearr{\begin{eqnarray} \lal}
\def\ear{\end{eqnarray}}
\def\earn{\nonumber \end{eqnarray}}

\def\nnn{\nonumber\\ \lal }

